\documentclass[conference]{IEEEtran}

\usepackage[T1]{fontenc}
\usepackage{microtype}
\usepackage{graphicx}
\usepackage{booktabs}
\usepackage{amsmath,amssymb}
\usepackage{subcaption}
\usepackage{siunitx}
\usepackage[hidelinks,bookmarks=false]{hyperref}
\usepackage{orcidlink}
\usepackage{cleveref}
\usepackage{cite}
\usepackage[font=small]{caption}
\usepackage{algorithm}
\usepackage{algpseudocode}
\usepackage{enumitem}

\makeatletter
\patchcmd{\algorithmic}{\addvspace{1ex}}{\addvspace{0.3ex}}{}{}
\makeatother

\crefname{section}{Section}{Sections}
\crefname{subsection}{Section}{Sections}
\crefname{figure}{Fig.}{Figs.}
\crefname{table}{Table}{Tables}
\crefname{equation}{Equation}{Equations}

\begin{document}

\title{CSI Simulation: Why Additive Noise Fails and \\How to Fix It}

\author{
\IEEEauthorblockN{Aymen Bouferroum}
\IEEEauthorblockA{\textit{Inria Lille-Nord Europe}\\
Lille, France\\
\orcidlink{0009-0002-9133-3152}\,0009-0002-9133-3152}
\and
\IEEEauthorblockN{Ildi Alla}
\IEEEauthorblockA{\textit{University of Luxembourg}\\
Luxembourg, Luxembourg\\
\orcidlink{0009-0008-6290-6998}\,0009-0008-6290-6998}
\and
\IEEEauthorblockN{Vincent Lenders}
\IEEEauthorblockA{\textit{University of Luxembourg}\\
Luxembourg, Luxembourg\\
\orcidlink{0000-0002-2289-3722}\,0000-0002-2289-3722}
\and
\IEEEauthorblockN{Valeria Loscri}
\IEEEauthorblockA{\textit{Inria Lille-Nord Europe}\\
Lille, France\\
\orcidlink{0000-0003-2558-1801}\,0000-0003-2558-1801}
}

\maketitle

\begin{abstract}
Channel State Information (CSI) has become a widely used wireless channel sensing modality for applications such as indoor localization, activity recognition, and respiration monitoring. Because collecting labeled data under every target condition is impractical, training CSI-based models often relies on simulated data produced by adding noise or perturbations to recorded channel estimates, most commonly additive white Gaussian noise (AWGN). This practice assumes that the receiver chain between the antenna and the channel estimator is \textit{linear and gain-invariant}. We test this assumption empirically using RF jamming as a controlled perturbation on 6 commodity receivers across 2 indoor environments. The assumption does not hold. Automatic gain control compresses the channel estimate multiplicatively before digitization, producing amplitude distributions that no additive noise variance can reproduce. To close the resulting fidelity gap, we propose \textbf{M\textsubscript{QTC}}, a measurement-calibrated model that learns the per-subcarrier distribution transformation through quantile mapping, temporal filtering, and copula-based cross-subcarrier reordering. M\textsubscript{QTC} reduces amplitude error 8-fold and closes 89\% of the aggregate fidelity gap across four complementary dimensions. The improvement transfers directly to downstream tasks, where 5 classifiers from different families trained on M\textsubscript{QTC}-simulated data recover 93\% of real-data jamming detection performance, while AWGN-trained classifiers remain near chance.
\end{abstract}


\vspace{-0.1cm}

\section{Introduction}
\label{sec:intro}

Channel State Information (CSI) captures the complex-valued frequency response that an orthogonal frequency-division multiplexing (OFDM) receiver estimates on every decoded frame as part of channel equalization. Because these per-subcarrier estimates capture fine-grained amplitude and phase distortions imposed by the propagation environment, CSI has become a general-purpose sensing modality for indoor localization, human activity recognition, gesture detection, and respiration monitoring~\cite{wang2026generalizability,alla2025sec5gloc}. CSI extraction is now supported across commodity Wi-Fi chipsets~\cite{halperin2011tool,gringoli2019freecsi,espressif2024espcsi}, enabling broad deployment for sensing research.

Many of these applications depend on simulated or synthetically augmented CSI for model training. CrossSense generates synthetic CSI from a single measurement set to enable cross-site sensing~\cite{zhang2018crosssense}. Noise-based perturbation of recorded CSI is a standard augmentation strategy in gesture recognition~\cite{wang2022airfi}, activity recognition~\cite{strohmayer2024augmentation}, indoor localization~\cite{serbetci2023channel}, and 5G positioning~\cite{gao2022toward}. These approaches inherit, explicitly or implicitly, the textbook additive white Gaussian noise (AWGN) model~\cite{goldsmith2005wireless}. Interference is added to a clean signal in the complex domain, where the model is well founded for raw baseband signals. In-phase and quadrature (I/Q)-level tools such as JamRF~\cite{jamrf2022github} generate interference waveforms along the same lines. The \textit{critical assumption} is that what holds for raw baseband signals also holds for CSI after receiver processing. This assumption has not been empirically tested. Generative models bypass the additive formulation by learning CSI distributions directly~\cite{bhat2024csi4free,wang2025genai}, but require large training sets and have not been validated against controlled measurements. A recent study on Wi-Fi sensing augmentation concluded that no prior work had systematically explored radio data augmentation, and that existing approaches are ad-hoc~\cite{hou2024rfboost}.

Before a received signal becomes a channel estimate, it passes through automatic gain control (AGC), analog-to-digital conversion (ADC), a fast Fourier transform (FFT) stage, and channel estimation from known training symbols~\cite{rappaport2024wireless}. Each stage introduces nonlinear distortions that an additive model does not capture. When interference raises total received power, AGC reduces the analog front-end gain before digitization, compressing the entire channel estimate by a factor that depends on the instantaneous power level~\cite{ratnam2024optimal,wei2023rssicsi}. This compression is multiplicative and acts on signal and noise together, so the amplitude distribution reported by the receiver has a fundamentally different shape than what an additive model predicts. Baseband filtering adds further subcarrier-dependent distortions~\cite{jiang2022picoscenes}, and a controlled cable-coupled comparison of commercial receivers recently confirmed that AGC behavior is the dominant source of cross-device CSI variation~\cite{portner2026receiver}. The additive simulation model ignores all of these stages. Whether the resulting CSI distributions are realistic is therefore not a modeling detail but a testable empirical question.

RF jamming provides an ideal case study for this test. A jammer injects a controlled waveform of known power and bandwidth, producing a measurable perturbation at the CSI level that can be directly compared against simulation predictions. Because jamming raises total received power substantially, it forces a large AGC response and makes any deviation from the additive model clearly visible. Jamming also carries practical relevance. CSI-based interference detectors need realistic training data, yet collecting paired clean and jammed recordings requires controlled environments and authorization to transmit interference~\cite{xu2005feasibility,pirayesh2022jamming}. Recent empirical work confirmed that CSI carries exploitable jamming signatures~\cite{mykytyn2025dcoss}, and simulation-based jamming detectors have been proposed without validating the fidelity of the underlying simulation~\cite{panitsas2025jamshield}. In developing our own CSI-based jamming detection and classification pipeline~\cite{bouferroum2026citadel}, we observed that simulated CSI distributions diverge substantially from real receiver output, motivating the systematic investigation presented in this work. The \textit{sim-to-real gap} in RF machine learning is a recognized problem across localization~\cite{manukyan2025bridging} and signal classification~\cite{clark2021training}, but it has not been quantified at the CSI level.

The contributions of this work are as follows.
\begin{itemize}
\item Empirical falsification of the receiver-chain linearity assumption: the AWGN model produces a Wasserstein distance of 3.09 compared to 0.39 for the proposed model, an 8-fold gap, providing, to our knowledge, the first quantitative test at the CSI level.
\item A calibrated simulation pipeline, M\textsubscript{QTC} (Quantile-Temporal-Copula), that closes 89\% of the aggregate fidelity gap across amplitude, phase, temporal, and spectral dimensions, validated on three external public datasets beyond the jamming scenario. Ablation identifies copula-based cross-subcarrier reordering as the dominant mechanism (9-fold aggregate reduction).
\item Sim-to-real transfer validation: classifiers trained on M\textsubscript{QTC}-simulated data recover 93\% of the detection performance achieved with real training data (AUC 0.904 vs.\ 0.967) across 5 classifier families and 5 receivers, while AWGN-trained classifiers operate near random decision (AUC 0.522).
\end{itemize}

To our knowledge, the paired clean and jammed CSI recordings released with this work are the first such public dataset. The framework code, dataset, and evaluation scripts are available online.\footnote{\url{https://github.com/aymenbouferroum/sim2real-csi}}
\section{Background and related work}
\label{sec:background}

\subsection{CSI simulation practice}

The simulation approaches used throughout the CSI literature fall into three categories, none of which explicitly model how the receiver chain transforms the signal before the channel estimator reports it.

The first and most common category is \textit{additive noise injection}. Complex Gaussian noise is added directly to recorded CSI to simulate degraded conditions. Table~\ref{tab:noise-practice} catalogues representative work that follows this pattern across five application domains. The underlying model is straightforward. If the true channel is $H_k$ and independent Gaussian noise $N_k$ is added, the simulated output is $\hat{H}_k = H_k + N_k$. This formulation is correct for raw baseband signals, where the noise enters before any receiver processing. The implicit assumption, rarely stated explicitly, is that the same formulation holds after the receiver chain has processed the signal. Serbetci et al.~\cite{serbetci2023channel} observed that naive Gaussian injection produces worse localization performance than physics-motivated methods modeling oscillator drift and amplifier fluctuations, but did not identify the receiver chain as the underlying cause.

The second category operates at the \textit{I/Q level}~\cite{alla2026m2rf}. Tools such as JamRF~\cite{jamrf2022github} generate interference waveforms for software-defined radio (SDR) transmission, producing physically correct baseband signals that nevertheless skip the receiver entirely. The gap between I/Q-level and CSI-level simulation has not been measured. A signal that is received at the antenna may produce a very different channel estimate after passing through AGC, digitization, and the FFT.

The third category uses \textit{generative models}. Bhat et al.~\cite{bhat2024csi4free} applied generative adversarial networks (GANs) to synthesize mmWave CSI for pose classification, and Wang et al.~\cite{wang2025genai} used conditioned diffusion models for integrated sensing and communication (ISAC) networks. These approaches learn the data distribution directly but require large training sets and have not been validated against controlled measurements where ground truth is available. All three categories share a \textit{common limitation}. The simulator output has not yet been compared, distribution by distribution, to what a real receiver produces under identical conditions.

\begin{table}[t]
  \centering
  \caption{Representative work adding noise or perturbation directly to CSI. Whether the resulting distributions match real receiver output has not been tested.}
  \label{tab:noise-practice}
  \small
  \setlength{\tabcolsep}{2pt}
  \begin{tabular}{@{}lll@{}}
    \toprule
    Reference & Domain & Noise method \\
    \midrule
    Wang et al.~\cite{wang2022airfi} & Gesture recog. & Additive Gaussian \\
    Strohmayer et al.~\cite{strohmayer2024augmentation} & Activity recog. & Amplitude scaling \\
    Barahimi et al.~\cite{barahimi2024context} & Activity recog. & $\mathcal{N}(0, 0.1)$ to CSI \\
    Thariq Ahmed et al.~\cite{alhazbi2020dfwislr} & Sign language & AWGN to raw CSI \\
    Serbetci et al.~\cite{serbetci2023channel} & Localization & $\mathcal{CN}(0, P)$ at SNR \\
    Gao et al.~\cite{gao2022toward} & 5G positioning & Noise at 3 SNR levels \\
    Zhang et al.~\cite{zhang2018crosssense} & Cross-site sensing & Synthetic generation \\
    \bottomrule
  \end{tabular}
\end{table}

\vspace{-0.1cm}
\subsection{Receiver-chain effects on CSI}

The receiver hardware between the antenna and the channel estimator introduces systematic distortions that have been studied independently but not connected to simulation validity. Ratnam et al.~\cite{ratnam2024optimal} developed a mathematical model for the gain and phase errors that AGC, sampling clock offset, and carrier frequency offset introduce into Wi-Fi CSI, and showed that these errors must be compensated before CSI can be used reliably for sensing. Wei et al.~\cite{wei2023rssicsi} demonstrated that AGC-induced amplitude scaling explains the discrepancy between received signal strength indicator (RSSI) and CSI power measurements on commodity devices. Jiang et al.~\cite{jiang2022picoscenes} showed that baseband filters introduce subcarrier-dependent amplitude and phase distortions that are pervasive across Wi-Fi network interface card (NIC) architectures and proposed the PicoScenes platform to expose these effects. Most recently, Portner et al.~\cite{portner2026receiver} conducted a controlled cable-coupled comparison of commercial receivers and confirmed that AGC behavior, rather than channel propagation, is the dominant source of cross-device CSI variation. Cominelli et al.~\cite{cominelli2023exposing} demonstrated that CSI sensing performance depends strongly on the deployment environment, with generalization across environments remaining an open problem. These studies establish that the receiver chain matters for CSI applications, but none connect it to how CSI is simulated.

\vspace{-0.1cm}
\subsection{Jamming detection and the validation gap}

The standard jammer taxonomy (constant, deceptive, random, reactive) was established by Xu et al.~\cite{xu2005feasibility}, who detected attacks from aggregated link metrics. Subsequent work continued to operate on link-level or spectrum-level representations. The transition to CSI-level analysis is recent. Mykytyn et al.~\cite{mykytyn2025dcoss} provided the first CSI-level characterization of jamming effects across static and mobile networks, confirming that jamming produces \textit{detectable per-subcarrier signatures} in both amplitude and phase. Panitsas et al.~\cite{panitsas2025jamshield} argued that detectors evaluated on simulated jamming data may not generalize to real over-the-air conditions, but did not trace the gap to the receiver chain or propose a calibrated simulation model.

The \textit{sim-to-real gap} is increasingly recognized across wireless machine learning. Manukyan et al.~\cite{manukyan2025bridging} reported that RF localization models achieving 25\,m error on synthetic data degraded to 184\,m on real measurements. Clark et al.~\cite{clark2021training} showed that synthetically trained RF classifiers lose significant accuracy without channel-aware augmentation. These studies confirm that naive simulation is insufficient, but operate at different signal representations (raw I/Q, RSSI, spectrograms) rather than at the CSI level. Validating simulated channels against physical measurements is standard practice in propagation modeling~\cite{rappaport2024wireless}, yet such validation has not been performed at the CSI level, and the mechanism that breaks the additive assumption remains unidentified.
\vspace{-7pt}
\section{The linearity assumption}
\vspace{-5pt}

\label{sec:assumption}

CSI simulation rests on one core assumption: the receiver chain between antenna and channel estimator is \textit{linear and gain-invariant}. Under this assumption, adding noise at the CSI output is equivalent to adding it at the antenna input. We state the assumption formally, explain why it fails in practice, and derive the baseline model that embodies it.

An OFDM receiver estimates the channel from known training symbols $X_k$ by observing $Y_k = H_k X_k + N_k$ and computing:
\begin{equation}
  \hat{H}_k = \frac{Y_k}{X_k}.
  \label{eq:chanest}
\end{equation}
When wideband interference adds complex Gaussian noise $J_k$ on every subcarrier, the jammed estimate becomes
\begin{equation}
  \hat{H}_k = H_k + \frac{J_k}{X_k},
  \label{eq:jammed}
\end{equation}
and since the training symbols have approximately unit magnitude in 802.11 OFDM, $J_k / X_k$ remains complex Gaussian with unchanged variance. If the receiver chain preserves linearity and gain, \cref{eq:jammed} predicts that interference simply adds independent Gaussian samples to the channel estimate. The standard AWGN approach~\cite{goldsmith2005wireless} operationalizes this prediction by adding independent complex Gaussian noise to each subcarrier of a clean recording:
\begin{equation}
  \hat{H}^{(j)}_k[n] = H^{(c)}_k[n] + \mathcal{CN}(0,\;\hat{\sigma}^2),
  \label{eq:m1}
\end{equation}
where the noise variance $\hat{\sigma}^2$ is calibrated from the mean per-subcarrier power difference between jammed and clean recordings:
\begin{equation}
  \hat{\sigma}^2 = \frac{1}{K}\sum_{k=1}^{K}\Bigl(\mathbb{E}\bigl[|H^{(j)}_k|^2\bigr] - \mathbb{E}\bigl[|H^{(c)}_k|^2\bigr]\Bigr).
  \label{eq:m1_sigma}
\end{equation}
This single-parameter model, which we refer to as \textbf{M\textsubscript{1}} (AWGN), treats all subcarriers identically, ignores temporal structure, and assumes amplitude distributions shift additively. It is the principled baseline against which the receiver-chain-aware models are measured.

\begin{figure}[t]
  \centering
  \includegraphics[width=\columnwidth]{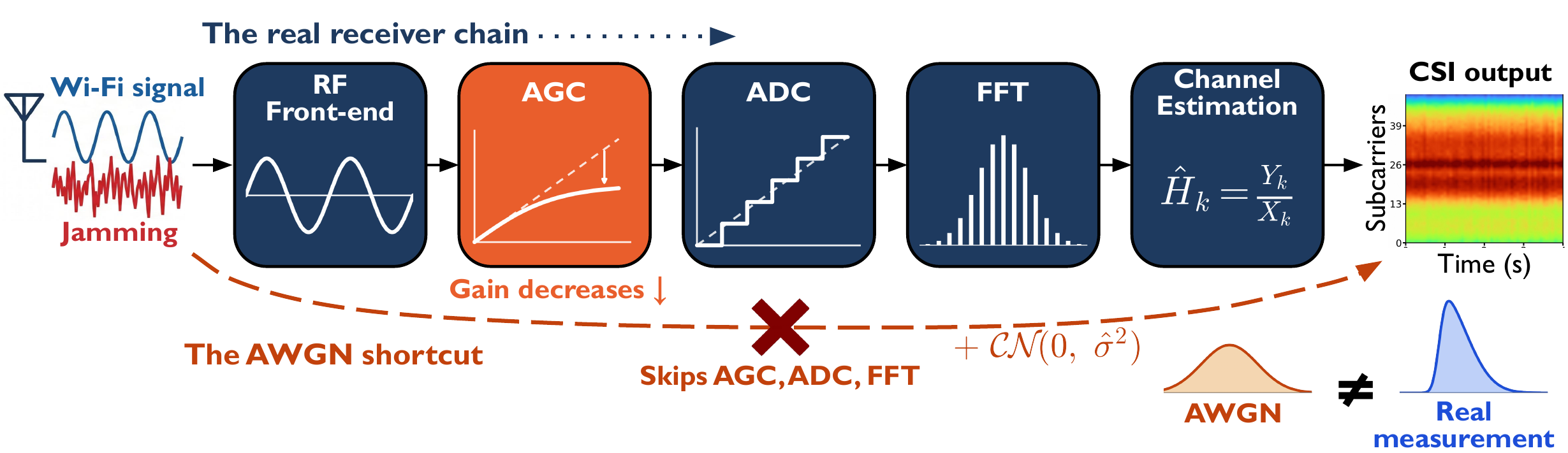}
  \caption{The receiver chain between the antenna and the CSI output. Real interference enters at the antenna and crosses AGC, digitization, and the FFT before reaching the channel estimator. M\textsubscript{1} adds noise directly at the output, skipping all three stages. The two paths coincide only if the chain is linear with fixed gain.}
  \label{fig:chain}
\end{figure}

The receiver, however, is not linear (\cref{fig:chain}). Three stages between the antenna and the channel estimator violate the assumption. \textit{First}, AGC adjusts the analog front-end gain based on total received power. When interference raises total power from $P_\text{clean}$ to $P_\text{clean} + P_\text{jam}$, AGC reduces the gain by a factor that depends on the power ratio, compressing the entire channel estimate before digitization~\cite{goldsmith2005wireless,ratnam2024optimal}. This gain reduction acts on signal and interference together. The jammed CSI amplitude changes far less than the injected power would suggest under a linear model, and the distributional shape is altered in a way that no scalar noise variance can reproduce. \textit{Second}, the analog-to-digital converter introduces quantization noise and, at high instantaneous power, clipping artifacts. At the moderate jamming power levels used in this study, quantization effects are secondary to AGC compression but become relevant at higher power levels where the AGC range is exceeded. \textit{Third}, subcarrier-dependent baseband filtering adds amplitude and phase distortions that vary across the frequency band~\cite{jiang2022picoscenes}. These three effects are present in every commodity Wi-Fi receiver, regardless of chipset~\cite{portner2026receiver}.

M\textsubscript{1} predicts that jammed amplitudes grow as the square root of added noise power, following a Rice or Rayleigh distribution shifted by independent Gaussian perturbations. A real receiver instead reports amplitudes compressed by AGC, with heavier tails from occasional gain-settling transients and cross-subcarrier correlations imposed by the wideband gain reduction. The assumption may hold approximately at very low interference levels, where the total received power changes negligibly and AGC gain remains nearly constant. As interference power increases, the multiplicative compression grows and the additive model becomes increasingly inaccurate. Whether this discrepancy matters for downstream applications is the empirical question addressed in \cref{sec:evaluation}. Two intermediate models that relax the assumption incrementally are introduced alongside the evaluation in \cref{sec:evaluation}.
\vspace{-0.31cm}
\section{M\textsubscript{QTC}: Quantile-Temporal-Copula model}
\label{sec:mqtc}
\vspace{-0.2cm}

\begin{figure}[t]
  \centering
  \includegraphics[width=\columnwidth]{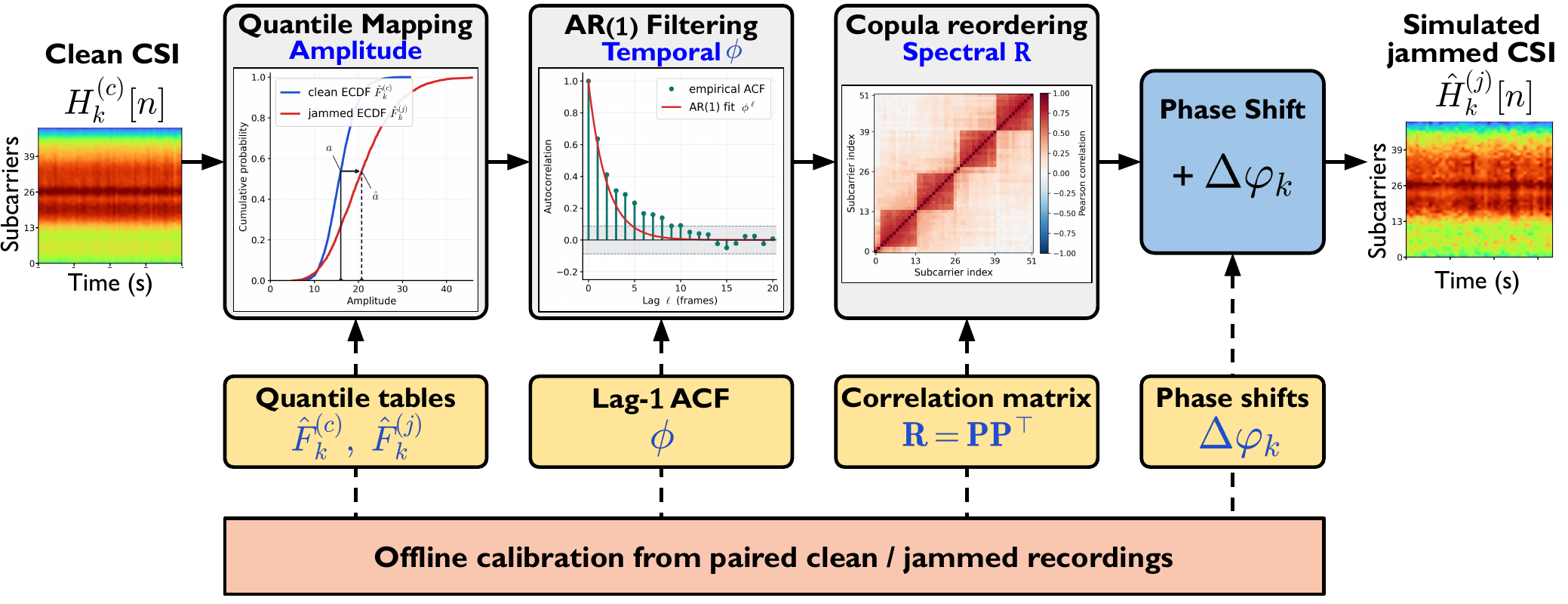}
  \caption{M\textsubscript{QTC} pipeline. Three calibrated stages: \textit{quantile mapping} (amplitude), \textit{AR(1) filtering} (temporal), and \textit{Iman-Conover copula reordering} (cross-subcarrier correlation).}
  \label{fig:mqtc-pipeline}
\end{figure}

M\textsubscript{QTC} addresses the three types of receiver-chain distortion identified in \cref{sec:assumption} through dedicated calibrated stages, as illustrated in \cref{fig:mqtc-pipeline}. It follows the same interface as M\textsubscript{1}: given paired clean and jammed recordings, calibrate parameters offline, then transform unseen clean CSI into simulated jammed CSI via $\hat{H}^{(j)}_k[n] = f_\theta\bigl(H^{(c)}_k[n]\bigr)$.

\indent \textbf{a) Stage 1: Per-subcarrier quantile mapping.}~
\textit{Quantile mapping} is a distribution-transfer technique widely used for bias correction in climate modeling~\cite{cannon2015quantile}. For each subcarrier $k \in \{1, \ldots, K\}$, we estimate the empirical cumulative distribution functions (ECDFs) $\hat{F}^{(c)}_k$ and $\hat{F}^{(j)}_k$ of the clean and jammed amplitude distributions from the calibration data, evaluated at $L$ equally spaced percentiles. During simulation, a clean amplitude value $a$ is mapped to the jammed domain by composition of the clean ECDF with the jammed quantile function:
\begin{equation}
  \hat{a}_k[n] = \bigl(\hat{F}^{(j)}_k\bigr)^{-1}\!\Bigl(\hat{F}^{(c)}_k\bigl(|H^{(c)}_k[n]|\bigr)\Bigr).
  \label{eq:qmap}
\end{equation}
Values at the tails are linearly extrapolated from the two nearest percentiles. This guarantees that the simulated marginal distribution for each subcarrier matches the jammed reference exactly, regardless of its shape. The mapping captures AGC compression, power broadening, and any other distortion present in the per-subcarrier amplitude histograms, without requiring an explicit hardware model. The storage cost is $2KL$ floating-point values (two ECDFs per subcarrier), and the per-frame mapping cost is $O(K \log L)$ via binary search.

\indent \textbf{b) Stage 2: AR(1) temporal filtering.}~
Quantile mapping produces correct marginals but no temporal structure beyond what the clean input carries. Real jammed CSI exhibits frame-to-frame autocorrelation that differs from the clean signal, because AGC gain adjustments operate on a time scale comparable to the frame interval. We model this with a first-order autoregressive process~\cite{box2015timeseries}, chosen because low frame rates typical of commodity CSI extraction are insufficient to resolve higher-order dynamics. We estimate the lag-1 autocorrelation $\phi$ from the median autocorrelation function (ACF) of amplitude residuals across subcarriers. Rather than adding temporally correlated noise on top of the quantile-mapped amplitudes (which would destroy the marginals), we use $\phi$ to filter the random scores that drive the rank reordering in Stage~3:
\begin{equation}
  s_k[n] = \phi \cdot s_k[n{-}1] + \sqrt{1-\phi^2}\cdot z_k[n],
  \label{eq:ar1}
\end{equation}
where $z_k[n]$ are Cholesky-correlated normal scores produced by the copula stage below (\cref{eq:reorder}) and the innovation scaling $\sqrt{1-\phi^2}$ preserves unit variance.

\indent \textbf{c) Stage 3: Iman-Conover copula reordering.}~After Stage~1, cross-subcarrier correlation still reflects the clean input rather than the wideband structure imposed by AGC. Since AGC reduces gain for the entire band simultaneously, independent per-subcarrier mapping destroys the correlated amplitude changes it creates. We restore the target correlation using the \textit{Iman-Conover method}~\cite{iman1982distribution}: from jammed calibration amplitudes, we estimate the $K \times K$ Pearson correlation matrix $\mathbf{R}$ and compute its Cholesky factor $\mathbf{P}$ (with diagonal regularization $\epsilon\mathbf{I}$). Standard normal samples $\mathbf{W} \in \mathbb{R}^{N \times K}$ are correlated via $\mathbf{Z} = \mathbf{W}\mathbf{P}^T$ and filtered through the AR(1) process to produce rank scores $\mathbf{S}$, which reorder the quantile-mapped amplitudes:
\begin{equation}
  \tilde{a}_k[n] = \operatorname{sort}\bigl(\hat{a}_k\bigr)\!\Bigl[\operatorname{rank}\bigl(s_k[n]\bigr)\Bigr].
  \label{eq:reorder}
\end{equation}
Because this step only permutes existing values (the sorted amplitudes are simply rearranged), it preserves the per-subcarrier marginals exactly while imposing the target cross-subcarrier dependence. The Cholesky decomposition is $O(K^3)$ but is computed once offline.

\indent \textbf{d) Phase reconstruction and output.}~
On commodity Wi-Fi receivers, phase estimation is typically dominated by hardware noise, limiting the benefit of complex phase models. We apply a per-subcarrier circular mean phase shift $\Delta\varphi_k$ estimated from the calibration data:
\begin{equation}
  \Delta\varphi_k = \arg\!\Biggl(\frac{1}{N}\sum_{n=1}^{N} \exp\Bigl(j\bigl(\angle H^{(j)}_k[n] - \angle H^{(c)}_k[n]\bigr)\Bigr)\Biggr).
  \label{eq:phaseshift}
\end{equation}
The final simulated jammed CSI is $\hat{H}^{(j)}_k[n] = \tilde{a}_k[n] \cdot \exp\bigl(j(\angle H^{(c)}_k[n] + \Delta\varphi_k)\bigr)$. More sophisticated phase models were not pursued because the circular-mean shift already matches the observed phase variation within measurement noise. \Cref{alg:mqtc} summarizes the online simulation procedure.

\vspace{-0.3cm}
\begin{figure}[H]
\begin{algorithm}[H]
\small
\caption{M\textsubscript{QTC} Online Simulation}
\label{alg:mqtc}
\begin{algorithmic}[1]
\Statex \textbf{Input:} Clean CSI $\{H^{(c)}_k[n]\}$; ECDFs $\hat{F}^{(c)}_k, \hat{F}^{(j)}_k$; Cholesky $\mathbf{P}$; AR coeff.\ $\phi$; phase shifts $\Delta\varphi_k$
\Statex \textbf{Output:} Simulated jammed CSI $\{\hat{H}^{(j)}_k[n]\}$
\For{each subcarrier $k = 1, \ldots, K$}
    \State $\hat{a}_k[n] \leftarrow (\hat{F}^{(j)}_k)^{-1}\!\bigl(\hat{F}^{(c)}_k(|H^{(c)}_k[n]|)\bigr)$ \Comment{Quantile map}
\EndFor
\State Draw $\mathbf{W} \sim \mathcal{N}(\mathbf{0}, \mathbf{I})$, size $N \times K$
\State $\mathbf{Z} \leftarrow \mathbf{W}\mathbf{P}^T$ \Comment{Induce cross-subcarrier correlation}
\For{each subcarrier $k = 1, \ldots, K$}
    \State $s_k[n] \leftarrow \phi \cdot s_k[n{-}1] + \sqrt{1{-}\phi^2} \cdot z_k[n]$ \Comment{AR(1) filter}
    \State $\tilde{a}_k[n] \leftarrow \text{sort}(\hat{a}_k)[\text{rank}(s_k[n])]$ \Comment{Copula reorder}
\EndFor
\State $\hat{H}^{(j)}_k[n] \leftarrow \tilde{a}_k[n] \cdot \exp\bigl(j(\angle H^{(c)}_k[n] + \Delta\varphi_k)\bigr)$ \Comment{Reconstruct}
\end{algorithmic}
\end{algorithm}
\end{figure}
\vspace{-0.3cm}

\indent \textbf{e) Computational cost.}~
Online simulation is $O(NK^2)$, dominated by the matrix multiply $\mathbf{Z} = \mathbf{W}\mathbf{P}^T$. Copula reordering adds $O(KN\log N)$ and quantile mapping $O(KN\log L)$, subdominant since $K$ exceeds both $\log N$ and $\log L$ for practical traces. M\textsubscript{1} runs in $O(NK)$, so M\textsubscript{QTC} adds a factor-$K$ overhead. Offline calibration is also $O(NK^2)$ from the correlation matrix, plus $O(K^3)$ Cholesky and $O(KN\log N)$ quantile estimation. Storage is $2KL + K^2 + K + 1$ floats.
\section{Experimental setup}
\label{sec:setup}

\subsection{Fidelity metrics}

A simulation model $f_\theta$ transforms clean CSI $H^{(c)}_k[n]$ into simulated jammed CSI $\hat{H}^{(j)}_k[n] = f_\theta(H^{(c)}_k[n])$. We measure fidelity as the mean distance across four dimensions:
\begin{equation}
  \mathcal{F}(\hat{H}^{(j)}, H^{(j)}) = \tfrac{1}{4}\sum_{d \in \mathcal{D}} F_d,
  \label{eq:fidelity}
\end{equation}
where $\mathcal{D}$ = \{amplitude, phase, temporal, spectral\}. The \textit{amplitude metric} uses the first Wasserstein distance~\cite{rubner2000emd} between per-subcarrier magnitude distributions, averaged across all $K$ subcarriers:
\begin{equation}
  F_{\text{ampl}} = \frac{1}{K}\sum_{k=1}^{K} W_1\!\bigl(|\hat{H}^{(j)}_k|,\; |H^{(j)}_k|\bigr).
  \label{eq:fampl}
\end{equation}
The \textit{phase metric} uses the absolute difference in circular variance~\cite{mardia2000directional}, where $V(\theta) = 1 - \bigl|\frac{1}{N}\sum_n e^{j\theta[n]}\bigr|$:
\begin{equation}
  F_{\text{phase}} = \frac{1}{K}\sum_{k=1}^{K} \bigl|V(\angle \hat{H}^{(j)}_k) - V(\angle H^{(j)}_k)\bigr|.
  \label{eq:fphase}
\end{equation}
The \textit{temporal metric} uses the $\ell_2$ norm of the ACF difference over $L_{\max}$ lags:
\begin{equation}
  F_{\text{temp}} = \frac{1}{K}\sum_{k=1}^{K} \Bigl\|\text{ACF}_{L_{\max}}\!\bigl(|\hat{H}^{(j)}_k|\bigr) - \text{ACF}_{L_{\max}}\!\bigl(|H^{(j)}_k|\bigr)\Bigr\|_2.
  \label{eq:ftemp}
\end{equation}
The \textit{spectral metric} uses the Frobenius norm of the difference between the $K \times K$ cross-subcarrier Pearson correlation matrices:
\begin{equation}
  F_{\text{spec}} = \bigl\|\mathbf{R}(\hat{H}^{(j)}) - \mathbf{R}(H^{(j)})\bigr\|_F.
  \label{eq:fspec}
\end{equation}
All four dimensions receive equal weight. Their natural scales differ substantially, so we report per-dimension scores alongside the aggregate throughout. Equal weighting avoids introducing arbitrary preferences across dimensions that have no common unit. The aggregate serves as a compact summary rather than the sole evaluation. The per-dimension breakdown in every table allows readers to assess each dimension independently and reweight according to their application.

\vspace{-0.1cm}
\subsection{Testbed and data collection}

\begin{figure}[t]
  \centering
  \includegraphics[width=0.9\columnwidth]{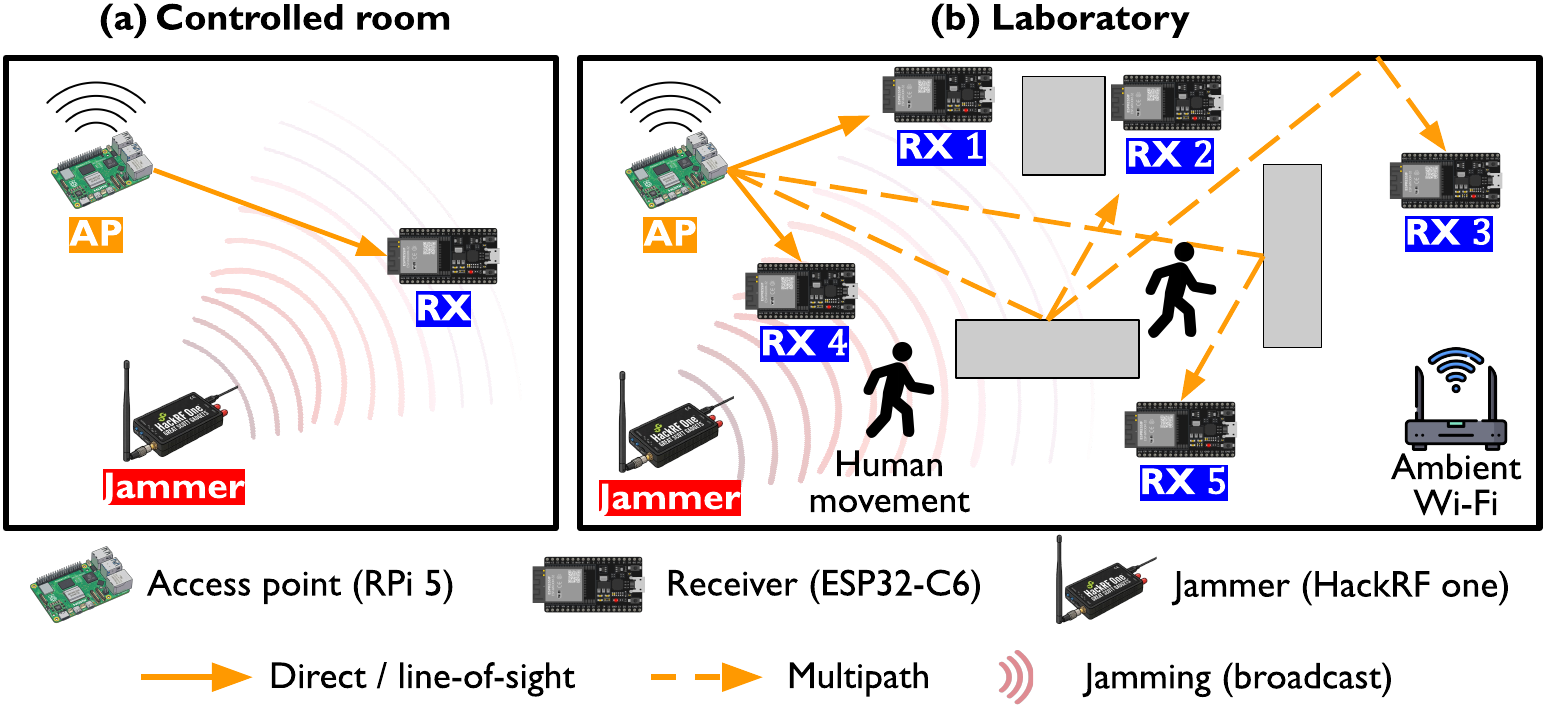}
  \caption{Experimental testbed. (a)~Controlled room: single ESP32-C6, Raspberry~Pi~5 AP, and HackRF One in fixed positions. (b)~Laboratory: 5 ESP32-C6 receivers in an active environment with ambient Wi-Fi and multipath.}
  \label{fig:testbed}
\end{figure}

Both environments share the same hardware (\cref{fig:testbed}). A Raspberry Pi~5 access point (802.11n HT20, channel~13), one or more ESP32-C6 receivers extracting $K{=}52$ subcarrier CSI estimates at roughly 3~fps via MQTT, and a HackRF One transmitting pre-generated jamming waveforms. The \textit{controlled room} has a single receiver in fixed line-of-sight positions with minimal ambient Wi-Fi. Each recording comprises approximately 20~minutes of clean CSI followed by 15~minutes of jammed CSI. The \textit{laboratory} deploys 5 receivers simultaneously at different positions and orientations in an active research environment with neighboring Wi-Fi networks and human movement. Laboratory recordings use interleaved 3-minute clean/jammed blocks to reduce drift. Constant Gaussian jamming was recorded at 5, 10, 15, and 20~dB intermediate frequency (IF) gain in the controlled room, and at 10, 15, and 20~dB in the laboratory. Within-scenario validation used 70/30 window-level train/test splits over windows of 32 consecutive frames, repeated over 5 random seeds. The controlled room isolates simulation fidelity from environmental confounds, while the lab tests whether findings generalize across devices and propagation conditions.

\vspace{-0.1cm}
\section{Evaluation}
\label{sec:evaluation}

We evaluate the linearity assumption (\cref{sec:assumption}) and M\textsubscript{QTC} (\cref{sec:mqtc}) using the testbed and metrics of \cref{sec:setup}.

\vspace{-0.1cm}
\subsection{Within-scenario fidelity}
\label{sec:assumption-fails}

To isolate which receiver-chain corrections matter, we define two intermediate models alongside M\textsubscript{1}. \textit{M\textsubscript{2} (Scaling)} scales each subcarrier by its jammed-to-clean power ratio, $\alpha_k = \bigl(\mathbb{E}[|H^{(j)}_k|^2]/\mathbb{E}[|H^{(c)}_k|^2]\bigr)^{1/2}$, and adds Gaussian noise with cross-subcarrier correlation estimated from the residuals. This model captures the correct mean power per subcarrier but still assumes the amplitude distribution retains a Gaussian shape after scaling. \textit{M\textsubscript{3} (Scaling+Corr)} adds Iman-Conover rank reordering on top of M\textsubscript{2}, matching the jammed cross-subcarrier correlation while preserving M\textsubscript{2}'s per-subcarrier marginals exactly.

\begin{table}[t]
  \centering
  \caption{Within-scenario fidelity at 20~dB, controlled room (mean$\pm$std, 5 seeds). Lower is better; phase is tied at this precision.}
  \label{tab:within}
  \small
  \setlength{\tabcolsep}{3.5pt}
  \begin{tabular}{lrrrrr}
    \toprule
    Model & Ampl. & Phase & Temp. & Spect. & Aggr. \\
    \midrule
    M\textsubscript{1} (AWGN)   & 3.09   & 0.02   & 0.53   & 19.84   & 5.87$\pm$0.06 \\
    M\textsubscript{2} (Scaling)  & 0.82   & 0.02   & 0.37   & 18.10   & 4.83$\pm$0.09 \\
    M\textsubscript{3} (Scaling+Corr)  & 0.82 & 0.02 & 0.58 & 1.48 & 0.73$\pm$0.04 \\
    M\textsubscript{QTC} & 0.39 & 0.02 & 0.49 & 1.61 & \textbf{0.63$\pm$0.07} \\
    \bottomrule
  \end{tabular}
\end{table}

\begin{figure}[t]
  \centering
  \includegraphics[width=\columnwidth]{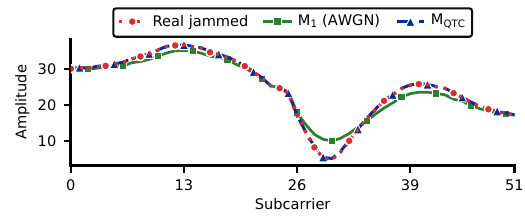}
  \caption{Mean amplitude across all 52 subcarriers at 20~dB. M\textsubscript{QTC} tracks the real jammed profile across the entire band, while M\textsubscript{1} diverges around the null subcarriers.}
  \label{fig:amplitude-profile}
\end{figure}

\Cref{tab:within} reports the results. The amplitude column directly tests the linearity assumption. M\textsubscript{1} produced a Wasserstein distance of 3.09, confirming that adding Gaussian noise at the measured power difference yields an amplitude distribution far from the one observed. \Cref{fig:amplitude-profile} makes this visible across the full band, where M\textsubscript{QTC} tracks the real jammed amplitude profile on all 52 subcarriers while M\textsubscript{1} diverges around the null subcarrier region. No scalar noise variance can reproduce a distribution whose shape was altered by a multiplicative gain stage. This finding is consistent with Serbetci et al.~\cite{serbetci2023channel}, who showed that naive Gaussian injection on CSI produces worse localization than physics-motivated augmentation.

\begin{figure}[t]
  \centering
  \includegraphics[width=\columnwidth]{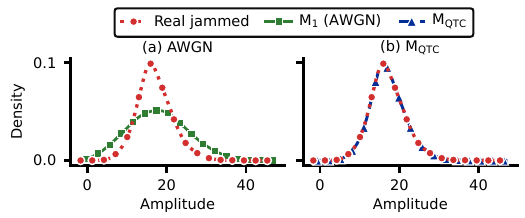}
  \caption{Amplitude density on subcarrier~26 at 20~dB. M\textsubscript{1} overshoots the real jammed distribution while M\textsubscript{QTC} closely tracks it.}
  \label{fig:kde}
\end{figure}

M\textsubscript{QTC} reduced the amplitude gap to 0.39, an 8-fold improvement, because quantile mapping learns the actual per-subcarrier distribution transformation rather than assuming additivity. \Cref{fig:kde} shows this in detail on subcarrier~26, where M\textsubscript{1} overshoots the real jammed distribution while M\textsubscript{QTC} closely tracks it. The aggregate score dropped from 5.87 to 0.63, closing 89\% of the gap.

Phase fidelity was identical across all models at approximately 0.02, because phase estimation on commodity receivers is dominated by hardware noise that masks the small phase perturbation introduced by jamming. M\textsubscript{2} reduced the aggregate only modestly because the mean power ratio is correct but the distributional shape remains wrong. M\textsubscript{3} achieved a larger reduction by restoring cross-subcarrier correlation, yet retains a 2-fold amplitude residual because it corrects the dependence structure but inherits M\textsubscript{2}'s Gaussian-shaped marginals, whereas M\textsubscript{QTC} replaces those marginals entirely through quantile mapping before reordering. In the laboratory (all 5 devices pooled), M\textsubscript{QTC} achieved an aggregate of 1.30 compared to 7.21 for M\textsubscript{1}, maintaining the same ordering.

\vspace{-0.1cm}
\subsection{Ablation}
\label{sec:ablation}

\Cref{tab:ablation} decomposes M\textsubscript{QTC} into four variants. All share the same amplitude score because quantile mapping is present in every configuration. The critical distinction is spectral. Q-only produced 21.15, \textit{worse} than M\textsubscript{1} (19.84), because independent per-subcarrier mapping destroys the cross-subcarrier correlation that the clean input carried. Correcting marginals alone is not sufficient when AGC imposes wideband dependence that independent processing erases.

\begin{table}[t]
  \centering
  \caption{Ablation of M\textsubscript{QTC} at 20~dB (mean$\pm$std, 5 seeds). Q = quantile mapping, T = AR(1), C = copula. Per-dimension scores: controlled room; aggregates: both environments.}
  \label{tab:ablation}
  \small
  \setlength{\tabcolsep}{3.5pt}
  \begin{tabular}{lrrr rr}
    \toprule
    Variant & Ampl. & Temp. & Spect. & Aggr. (ctrl) & Aggr. (lab) \\
    \midrule
    Q-only & 0.39 & 0.79 & 21.15 & 5.59$\pm$0.11 & 6.51$\pm$0.17 \\
    QT     & 0.39 & 0.50 & 20.30 & 5.30$\pm$0.05 & 7.97$\pm$0.08 \\
    QC     & 0.39 & 0.58 & 1.52 & 0.63$\pm$0.05 & 1.67$\pm$0.16 \\
    QTC    & 0.39 & 0.49 & 1.61 & 0.63$\pm$0.07 & \textbf{1.30$\pm$0.18} \\
    \bottomrule
  \end{tabular}
\end{table}

Adding copula reordering (QC) reduced the spectral score to 1.52 and the aggregate to 0.63, a 9-fold improvement over Q-only. Adding AR(1) temporal filtering (QT) improved the temporal dimension from 0.79 to 0.50 but did not address cross-subcarrier correlation. The full QTC achieved the same aggregate (0.63) as QC in this setting, because adding T improves the temporal dimension (QC 0.58 to QTC 0.49) but this gain is offset by a slight spectral shift. Copula reordering is the dominant mechanism across all conditions. The laboratory column in \cref{tab:ablation} confirms this in a second environment. QTC achieved the lowest aggregate (1.30$\pm$0.18), a 22\% improvement over QC (1.67$\pm$0.16), driven by better temporal fidelity (2.83 to 1.34). The temporal component's contribution is larger in the laboratory, where richer frame-to-frame dynamics are present. Across all seven tested conditions (four power levels in the controlled room, three in the laboratory), QTC achieved the lowest or tied-lowest aggregate, confirming that the full pipeline generalizes beyond a single operating point.

\vspace{-0.1cm}
\subsection{Per-device consistency}
\label{sec:perdevice}

\begin{figure}[t]
  \centering
  \includegraphics[width=\columnwidth]{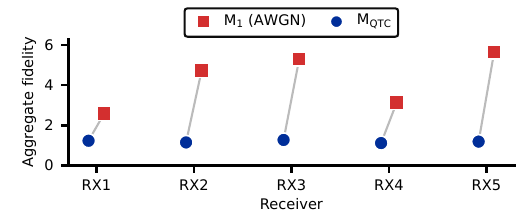}
  \caption{Per-device aggregate fidelity in the laboratory at 20~dB (5 seeds, lower is better).}
  \label{fig:perdevice-detection}
\end{figure}

\Cref{fig:perdevice-detection} reports per-device fidelity in the laboratory. The assumption failure is consistent. Every device shows large M\textsubscript{1} divergence (aggregate 2.58 to 5.67) and successful M\textsubscript{QTC} recovery (1.11 to 1.27), a 2.1- to 4.8-fold improvement. Per-device calibration produced lower scores than the merged-lab result (1.30), confirming that device-specific calibration captures variation that pooling averages out. In \textit{leave-one-device-out} evaluation, M\textsubscript{QTC}'s aggregate rose to 5.82, confirming that the transformation is position-specific. This is expected, as AGC compression depends on total received power, which varies with position and orientation, tying calibration to deployment geometry. The practical cost is a single 15-minute paired recording, after which unlimited realistic data can be generated from any clean recording without additional jammer hardware.

\subsection{Sim-to-real detection}
\label{sec:sim2real}

The downstream consequence of simulation fidelity becomes visible at moderate power. \Cref{tab:detection} evaluates at 10~dB IF gain, the lowest laboratory jammer power, where the detection task is most challenging. We used 5 classifiers spanning different families: logistic regression (LR), support vector machine (SVM) with RBF kernel, random forest (RF), a two-layer multilayer perceptron (MLP), and an autoencoder (AE)-based classifier (encoder-decoder with a classification head trained jointly on reconstruction and binary cross-entropy losses). Each classifier was trained on $K$-dimensional per-subcarrier mean amplitude features from simulated jammed data and tested on real jammed CSI (3 seeds per device, averaged across devices). An oracle trained on real jammed data served as the upper bound.

\begin{table}[t]
  \centering
  \caption{Sim-to-real detection AUC at 10~dB in the laboratory (per-device calibration, mean$\pm$std across 5 receivers, 3 seeds).}
  \label{tab:detection}
  \small
  \setlength{\tabcolsep}{3.5pt}
  \begin{tabular}{lrrrrr|r}
    \toprule
    Training data & LR & SVM & RF & MLP & AE & Mean \\
    \midrule
    Oracle (real) & 0.961 & 0.989 & 0.990 & 0.900 & 0.997 & 0.967$\pm$0.009 \\
    \midrule
    M\textsubscript{QTC} & \textbf{0.920} & \textbf{0.905} & \textbf{0.929} & \textbf{0.817} & \textbf{0.949} & \textbf{0.904$\pm$0.022} \\
    M\textsubscript{3} & 0.875 & 0.756 & 0.866 & 0.771 & 0.850 & 0.824$\pm$0.061 \\
    M\textsubscript{2} & 0.836 & 0.848 & 0.878 & 0.742 & 0.828 & 0.826$\pm$0.062 \\
    M\textsubscript{1} (AWGN) & 0.469 & 0.408 & 0.623 & 0.461 & 0.649 & 0.522$\pm$0.051 \\
    \bottomrule
  \end{tabular}
\end{table}

M\textsubscript{QTC} achieved the highest AUC for every classifier, with a mean of 0.904 vs.\ 0.522 for M\textsubscript{1} and 0.967 for the oracle, recovering 93\% of oracle performance. Three of five M\textsubscript{1}-trained classifiers fell \textit{below a random classifier}. The additive model overestimates the amplitude shift, so classifiers learn incorrect decision boundaries that invert when applied to real data where AGC has compressed the amplitude differences.

\subsection{External validation}
\label{sec:external}

The receiver-chain distortion that breaks the additive assumption is not specific to jamming. AGC compression acts on any change in total received power, whether caused by interference, human presence, or environment change. To test whether M\textsubscript{QTC}'s advantage generalizes, we compared it against four published CSI augmentation methods on three external public datasets spanning two hardware platforms. Wallhack1.8k~\cite{strohmayer2024augmentation} provides paired no-person and activity CSI from ESP32 receivers (52 subcarriers), SignFi~\cite{ma2019signfi} provides paired lab and home recordings from Intel~5300 receivers (30 subcarriers), and Widar~3.0~\cite{zheng2019widar} provides paired cross-room gesture recordings from Intel~5300 receivers (30 subcarriers). Each method was given its most generous calibrated form, and fidelity was measured as the amplitude Wasserstein distance on held-out test data (5 seeds).

\begin{table}[t]
  \centering
  \caption{Amplitude Wasserstein distance on three external public datasets (5 seeds, lower is better). Wallhack1.8k uses ESP32 (52 subcarriers), SignFi and Widar use Intel~5300 (30 subcarriers).}
  \label{tab:external}
  \small
  \setlength{\tabcolsep}{3.5pt}
  \begin{tabular}{lrrr}
    \toprule
    & Wallhack1.8k & SignFi & Widar~3.0 \\
    Method & (presence) & (lab$\to$home) & (room change) \\
    \midrule
    M\textsubscript{QTC}                              & \textbf{0.19} & \textbf{0.96} & \textbf{0.06} \\
    Gao et al.~\cite{gao2022toward}                   & 0.55          & 14.41         & 1.31          \\
    AirFi~\cite{wang2022airfi}                        & 0.37          & 13.83         & 1.30          \\
    Strohmayer et al.~\cite{strohmayer2024augmentation}& 0.51          & 8.28          & 0.34          \\
    Serbetci et al.~\cite{serbetci2023channel}         & 0.55          & 10.23         & 0.60          \\
    \bottomrule
  \end{tabular}
\end{table}

\Cref{tab:external} reports the results. M\textsubscript{QTC} achieved the lowest amplitude error on all three datasets, with a 2-fold improvement over the best baseline on Wallhack (0.19 vs.\ AirFi at 0.37), an 8.6-fold improvement on SignFi (0.96 vs.\ Strohmayer et al.\ at 8.28), and a 5.7-fold improvement on Widar (0.06 vs.\ Strohmayer et al.\ at 0.34). The advantage holds across both hardware platforms and all three types of CSI modification. All baseline methods apply additive or multiplicative perturbations that preserve the distributional shape, whereas M\textsubscript{QTC}'s quantile mapping learns the true per-subcarrier distribution regardless of its form.

\subsection{Computational cost}
\label{sec:cost}

For our configuration ($K{=}52$, $L{=}1000$ quantile percentiles), M\textsubscript{QTC} storage is 0.8\,MB. Simulating 1\,000 frames takes approximately 14\,ms on a single CPU core, roughly 6 times slower than M\textsubscript{1} but more than $10^4$ times faster than the 3\,fps collection rate. Calibration from 3\,000 frames completes in under one second. The overhead relative to M\textsubscript{1} is modest given the 8-fold fidelity improvement, and the entire calibrate-then-simulate cycle remains fast enough to generate unlimited realistic data once a single paired recording is available.
\section{Discussion}
\label{sec:discussion}

Because AGC operates upstream of the FFT, the distortion mechanism identified in this work is present in OFDM receivers across standards, suggesting that the assumption failure extends to 802.11be and 5G~NR, though validation on those platforms remains future work. Our experience developing a real-data CSI-based detection pipeline~\cite{bouferroum2026citadel} highlighted the gap between simulated and measured receiver output that motivated this investigation. Calibrated simulation could complement such systems by enabling rapid data generation for new deployment sites from a single paired recording.
The cost of this fidelity is position-specificity. M\textsubscript{QTC}'s richer parameterization captures the local AGC operating point and must be recalibrated at each deployment, consistent with the environment dependence reported for CSI sensing~\cite{cominelli2023exposing}. M\textsubscript{1}, by contrast, transfers with negligible change because its single parameter carries no position-specific information. The residual amplitude error shrinks further with larger calibration sets, reaching the 0.10 sampling-noise floor when calibration and evaluation use the same recording. For wider-bandwidth systems where $K$ exceeds 1\,000, the correlation matrix is naturally banded, so block-diagonal or sparse approximations can maintain tractability.
All experiments in this study used the ESP32-C6 chipset, though the external validation (\cref{sec:external}) provides indirect evidence of cross-platform applicability on Intel~5300 data. Platforms based on Intel~5300~\cite{halperin2011tool} or Nexmon-modified Broadcom~\cite{gringoli2019freecsi} may exhibit different distortion patterns, but Portner et al.~\cite{portner2026receiver} confirmed via cable-coupled measurements that AGC-induced amplitude scaling is the dominant cross-device CSI effect across Intel, Broadcom, and Espressif chipsets, so the assumption failure is structural, not chipset-specific.
\section{Conclusion}
\label{sec:conclusion}

This paper presented, to our knowledge, the first empirical validation of CSI-level simulation against real wireless measurements. Using paired clean and jammed recordings from 6 commodity Wi-Fi receivers across 2 environments, we showed that the standard AWGN model fails because the receiver chain transforms interference multiplicatively before the channel estimator reports it. The proposed M\textsubscript{QTC} pipeline closes 89\% of the fidelity gap through measurement-calibrated statistical transforms, with copula reordering as the dominant mechanism. Fidelity translates to utility. M\textsubscript{QTC}-trained classifiers recovered 93\% of oracle detection performance while AWGN-trained classifiers remained near random classifiers. The broader takeaway is that simulation models must be validated against the \textit{hardware they target}, not just against the channel model they implement.

\section*{Acknowledgment}
This work was supported by the French National Research Agency (ANR) under grant ANR-22-PEFT-0007 (NF-FITNESS project) as part of the France 2030 program, and by the Luxembourg National Research Fund under the PEARL project P24/IS/18423066/SCICT.

\bibliographystyle{IEEEtran}
\bibliography{references}

\end{document}